\documentstyle[aps,multicol,eqsecnum,epsf]{revtex}
\begin{document}
\draft
\title{Dry Friction due to Adsorbed Molecules} 
\author{C. Daly, J. Zhang and J. B. Sokoloff}
\address{Physics Department and Center for Interdisciplinary Research on
Complex Systems, Northeastern University, Boston, MA 02115.}
\date{\today }
\maketitle

\begin{abstract}
Using an adiabatic approximation method, which searches for Tomlinson 
model-like instabilities for a simple 
but still realistic model for two crystalline surfaces in the extremely light 
contact limit,  with mobile molecules present at 
the interface, sliding relative to each 
other, we are able to account for the virtually universal occurrence of "dry 
friction." The model makes important predictions for the dependence of friction 
on the strength of the interaction of each surface with the mobile molecules.
\end{abstract}

\begin{multicols}{2}
\narrowtext

      Muser and co-workers have argued that clean surfaces 
should not exhibit static friction [1], but 
the presence of mobile molecules (so called "third bodies) at the interface
can lead to static friction. This is a surprising result because one usually 
expects such lubricant molecules to reduce rather than enhance friction. On the 
other hand, if the mobile molecules are much more strongly attached to one 
surface than the other, they will act as randomly distributed pinning sites 
belonging to the surface to which they are strongly attached, and it was 
argued in Ref. 2 that 
molecular level random defects on the surface will not lead to static 
friction. Thus, an important ingredient in these 
molecules' leading to static friction is the relative strength of the 
interactions of the lubricant molecules with the two surfaces. 
We have done simple calculations which demonstrate that 
when the interaction of a molecule with the two 
surfaces is nearly of equal strength, the system exhibits multistability 
(i.e., the molecule can have two or more possible equilibrium 
positions for a given relative displacement of the two wells, one of which 
becomes unstable). This opens the exciting possibility that the relative 
strengths of the bonding to each of the surfaces of molecules trapped at 
an interface can be responsible for whether the molecules reduce or increase 
friction. It was argued by Caroli and co-workers [4] that without multistability 
there is no static or dry friction. There have been recent 
molecular dynamics studies of slow speed kinetic friction which relate 
their results to the mechanism of Ref. 4, both in one dimensional and 
two dimensional models[5]. The present work differs from Ref. 5 in that 
we have developed an adiabatic approximation method for locating 
Tomlinson-like potential instabilities which result in "dry friction" 
in the Muser-Robbins picture[1]. We feel that our adiabatic approximation 
method is more suited to the "dry friction" problem than molecular 
dynamics because it is better able to deal with the slow speed sliding 
limit.

      The model we have have studied consists of two rigid surfaces with 
a dilute concentration of particles trapped between them. To zeroth 
order, we neglect the particle-particle interactions. The 
surfaces are represented by two identical two 
dimensional periodic potentials, which are rotated relative to each 
other at an arbitrary angle, as this is the usual situation at an interface. 
We model the potential function acting on a mobile molecule due to each 
surface by the Steele potential[6] in the limit in which the molecule 
is not too close to either surface (compared to a lattice constant). For one 
surface (surf1), it is given by
 $$v_1(x,y)=v_0\sum_{\bf G} e^{i{\bf G}\cdot {\bf r}}=$$
$$v_0 \{2cos[(2\pi/a)x]cos[(2\pi/3^{1/2}a)y]]+
 cos[(4\pi/3^{1/2}a)y]\},  
 \eqno (1)$$
where the vectors {\bf G} denote the smallest reciprocal lattice vectors of 
a triangular lattice of lattice constant a and $v_0$ is the strength of 
the potential. This approximate potential is valid 
if the surfaces are just barely touching (but this is not contactless 
friction). We chose for the potential 
of the second surface (surf2),  the 
potential given in Eq. (1) rotated by $\phi$ and translated by 
$(\Delta x,\Delta y).$  Then this 
potential is given by $v_2 (x,y)=v_1 (x',y')$, 
where 
 $x'=(x+\Delta x)cos(\phi)+(y+\Delta y)sin(\phi)$ and
 $y'=-(x+\Delta x)sin(\phi)+(y+\Delta y)cos(\phi),$ where, 
 $\phi$ is the rotation angle, and the displacement parameters $\Delta x$ 
 and $\Delta y$ are given by:
 $\Delta x=s_0 cos(\theta)+bsin(\theta)$ and 
$\Delta y=s_0 sin(\theta)-bcos(\theta).$
Here, s0 = vt where 
v is the velocity of sliding of surf2 relative to surf1 along a direction 
making an angle $\theta$ with the x-axis. The minimum at the origin of surf2 
is moving along 
a path displaced a distance b, the distance of closest approach, normal 
to the path with respect to the minimum 
at the origin of surf1. 

Since we are neglecting intermolecular interaction, we study a single molecule 
placed at random within the Wigner Seitz unit cell of surf1 containing the 
origin, for an arbitrary value of b. We assume that each molecule will move 
to the nearest minimum of $v_1+v_2$. The resulting potential minimum reaches 
its smallest value when the two 
surfaces have slid until the two minima are at their distance of closest 
approach b. Therefore, the resulting potential minimum 
can only become unstable and disappear after this point, since before 
it the minimum is getting 
deeper. Thus we need only begin our search for instabilities 
for wells that are at their distance of closest approach. 
 Because this potential is a function of 
time, the existence of of these minima is time dependent. As the minimum 
disappears, the particle will drop to another potential minimum of lower 
energy, resulting in a gain of kinetic energy, which is assumed to get 
quickly transferred to phonons and electronic excitations of the surfaces. 
This is the mechanism for frictional energy dissipation. We have studied Eq. (1) 
using this method, but it is equally applicable to any two periodic or 
disordered potentials, representing the two surfaces. 

In order to locate minima, and to track their positions and 
and stability as our surfaces slide past one another, we first 
place a particle at a random position at the interface and use a 
Montecarlo routine to move it to the nearest potential minimum. 
In order to predict where the minimum will move during sliding, we use the fact 
that the force on a particle at the potential minimum $(x_0 (t),y_0 (t))$ 
remains identically zero for all time in the adiabatic approximation 
to find velocity at which the minimum is moving. Then, we have 
$${d\over dt}{\partial v\over \partial x}(x_0 (t),y_0 (t),t)=
{\partial^2 v\over \partial x^2}\vert_0 {dx\over dt}+
{\partial^2 v\over \partial x\partial y}\vert_0 {dy\over dt}+
{\partial^2 v\over \partial x\partial t}\vert_0=0, \eqno (2a)$$
$${d\over dt}{\partial v\over \partial y}(x_0 (t),y_0 (t),t)=
{\partial^2 v\over \partial y^2}\vert_0 {dy\over dt}+
{\partial^2 v\over \partial x\partial y}\vert_0 {dx\over dt}+
{\partial^2 v\over \partial y\partial t}\vert_0=0. \eqno (2b)$$
We then solve the above equations for the instantaneous velocities of the 
minimum as the surfaces slide as follows:
$${dx\over dt}={1\over D_0}{\partial^2 v\over \partial x\partial y}\vert_0
{\partial^2 v\over \partial y\partial t}\vert_0-
{\partial^2 v\over \partial y^2}\vert_0 {\partial^2 v\over \partial x\partial t
}\vert_0, \eqno (3a)$$
$${dy\over dt}={1\over D_0}{\partial^2 v\over \partial x\partial y}\vert_0
{\partial^2 v\over \partial x\partial t}\vert_0-
{\partial^2 v\over \partial x^2}\vert_0 {\partial^2 v\over 
\partial y\partial t}\vert_0. \eqno (3b)$$
Multiplying the velocity by the time step gives us the approximate new 
position of the minimum after sliding. The term in both dominators above,
which we have designated as $D_0$, is given by
$$D_0={\partial^2 v\over\partial x^2}_{0}{\partial^2 v\over \partial y^2}_{0}-
\vert {\partial^2 v\over \partial x\partial y}\vert_{0}^2.
\eqno (4)$$
It plays a critical role in our algorithm. It is 
known as the Gaussian curvature (for extrema points). When $D_0 = 0$ 
an instability occurs.  Furthermore, Eqs. 3 depend on the 
inverse of $D_0$. For this reason, the time step between successive relative 
displacements as the surfaces slide must be scaled by $D_0$ as we 
approach a minimum.

The 2nd order 
Taylor series expansion of the potential, assumed to be with respect to the 
location of the nearest minimum, 
$$v(x,y)=v(x_0,y_0)+{\partial v\over\partial x}\vert_0\Delta x+
{\partial v\over\partial y}\vert_0\Delta y$$
$$+(1/2){\partial^2 v\over\partial x^2}\vert_0\Delta x^2+
(1/2){\partial^2 v\over\partial x^2}\vert_0\Delta x^2+
(1/2){\partial^2 v\over\partial x\partial y}\vert_0\Delta x\Delta y, \eqno (5)$$
is now used to determine more accuratedly the location of the new minimum. 
The first order derivatives vanish, since we assume that we are expanding 
about the true minimum. The second order derivatives can, to 2nd order, be 
replaced by the second order derivatives at the present position of the 
particle, provided we are close to the actual minimum.  The quantities 
$\Delta x =(x_0-x_{pp})$ and $\Delta y=(y_0-y_{pp})$  are then the 
approximate distances, along the x and y directions, between the particle's 
present position (pp), and where the actual minimum is. In order to use the 
force 
components felt by the particle at it's present location to find $\Delta x$ 
and $\Delta y$, we 
differentiate the above 2nd order approximation with respect to both x and y, 
obtaining an approximation for the force components near the true minimum. 
$${\partial v\over\partial x}={\partial^2 v\over\partial x^2}\vert_{pp}
\Delta x+
{\partial^2 v\over\partial x\partial y}\vert_{pp}\Delta y, \eqno (6a)$$
$${\partial v\over\partial y}={\partial^2 v\over\partial y^2}\vert_{pp}
\Delta y+
{\partial^2 v\over\partial x\partial y}\vert_{pp}\Delta x. \eqno (6b)$$
Eqs. (6a) and (6b) are solved for $\Delta x$ and $\Delta y$ to give
$$\Delta x={1\over D}{\partial^2 v\over\partial y^2}\vert_{pp}
{\partial v\over\partial x}\vert_{pp}-
{\partial^2 v\over\partial x\partial y}\vert_{pp}{\partial v\over \partial y},
\eqno (7a)$$
 $$\Delta y={1\over D}{\partial^2 v\over\partial x^2}\vert_{pp}
{\partial v\over\partial x}\vert_{pp}-
{\partial^2 v\over\partial x\partial y}\vert_{pp}{\partial v\over \partial x},
\eqno (7b)$$
where D is the quantity given in Eq. (4) but evaluated at the point 
$(x_{pp},y_{pp})$. The derivatives in Eq. (7) are found from the potential 
$v_1+v_2$ defined in Eq. (1) and in the discussion under it. 
If the particle is close to the minimum, this procedure converges very quickly 
to the true minimum. How quickly it converges, however, is dependent on the 
size of the quantity D. 

 In our algorithm, we compute $D_0$. The second derivatives of the 
 potential form a two dimensional 2nd rank tensor, which can be 
diagonalized for appropriate orientation of the coordinate axes. $D_0$ 
is equal to the product of these diagonal elements. The xx 
component defines a parabola along the x-direction, and the yy component 
defines another along the y-direction. 
If both components are positive, one 
has a minimum, if both are negative, a maximum, and if one is positive and 
one is negative, then one has an instability, if the third order term in 
the Taylor series of Eq. (5) is nonzero and a minimum otherwise. 
When the minimum first becomes unstable, one of the eigenvalues, and hence 
$D_0$, goes to zero, we may have an instability. A typical instability is 
illustrated in Fig. 1, which shows a potential minimum which has become 
unstable, in the sense that one wall of the well minimum has disappeared, 
allowing a particle located in this minimum to flow into a neighboring 
minimum, which is also shown in this figure. 

\begin{figure}
\centerline{
\vbox{ \hbox{\epsfxsize=7.0cm \epsfbox{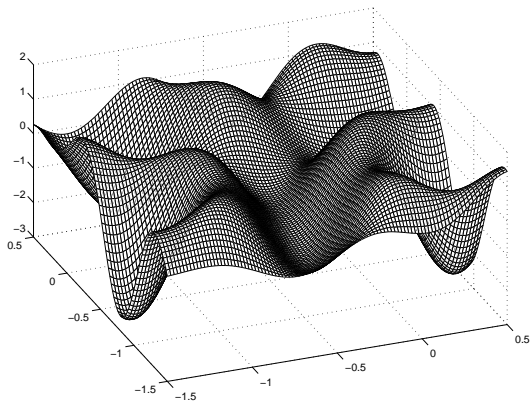}}
       \vspace*{1.0cm}
        }
}
\caption{An unstable potential minimum and a lower energy stable minimum are 
shown. The x and y axes are in units of a and the potential is in units of 
$v_0$.}
\label{Fig1}
\end{figure}

Our method allows us to track the position of a minimum 
until it becomes unstable, at which point we can locate the new minimum into 
which an unseated particle will next fall into. 
This allows us to calculate 
the drop in potential energy that such a particle would undergo, that we 
associate with the energy loss due to friction. The total frictional 
energy loss between our two surfaces would then be the sum of the energy drop 
for each particle every time it experiences an instability. Instead of doing a 
full simulation of many particles at an interface, which would be highly time 
consuming in the slow speed sliding limit, we have chosen to examine the 
motion in the adiabatic approximation of a single particle for various rotation 
angles $\phi$ and angles of sliding $\theta$. Then, 
a simple average is taken over the possible energy drops that occur for the 
various instabilities, in order to find the average energy 
loss between two surfaces, as a result of their sliding motion, for any 
number of particles. An absolute  minimum of the 
total potential is a result of the coalescence of two minima, one from each 
surface. Because of the periodicity of the surfaces, as these two minima 
slide past one another (for fixed values of 
$\phi$ and $\theta$), only one parameter is needed to describe the 
behaviors for the resulting potential minimum, the distance of 
closest approach b defined above. 
Because of this, we can examine all possible 
behaviors of a minimum of the total potential for fixed values of $\phi$ and 
$\theta$ by considering the behavior of the minimum that results from the 
overlap of the two 
central minima as a function of the parameter b. The results will give us all 
possible instabilities a single particle may undergo anywhere on the surface 
for a 
given rotation angle $\phi$ and angle of sliding $\theta$. From these results 
we can determine 
the average energy lost per particle per instability, and from 
this we 
can estimate the average frictional force between the two surfaces. Our results for 
one value of $\theta$ and $\phi$ are illustrated in the first two columns of table 1. 
Runs were made for all values of $0<b/a<0.5$ with a spacing of 0.02. Values of b/a 
for wich no instabilities were found are not listed in the table. 

In order to estimate the force of friction, we first find $<\Delta E>$, the mean 
value of the energy drop in an  instability for each value of $\theta$ and $\phi$. 
For example, for the values given in table 1, we obtain $<\Delta E>=0.0667v_0$. The 
mean value of the force of friction is given by the $<\Delta E>/<\Delta x>$, were 
$<\Delta x>$ denotes the mean distance that the two surfaces must be slid in order to find 
an instability. Since $<\Delta x>$ is of the order of a lattice spacing, which is of the 
order of $3\times 10^{-8}cm$, and since the potential strength $V_0$ is of the order of 
0.01980 eV[6,7], we obtain a force of friction per molecule at the interface of the order 
of $7.044\times 10^{-8} dyn$ for  $\theta=0.327$ rad and $\phi=0.1309$ rad. The values 
of $<\Delta E>$ for other values of $\theta$ and $\phi$ that we considered 
were of similar magnitude. If a unit cell area of a surface is of the order of 
$10^{-15}cm^2$ and there is a concentration of molecules (i.e. the number of molecules 
per unit cell) of 0.01, we obtain a frictional stress (i.e., the force of friction 
per $cm^2$ of contact area) of $7.044\times 10^5 dyn/cm^2$. Then an interface of total 
area $1cm^2$ with an area of contact (at asperities) which is 2 percent of this value, 
will exhibit a force of friction of $0.02cm^2$ times the frictional stress, or about 
$10^4 dyn$ or about 0.1 N, which is a reasonable value.

We have repeated our procedure for the case in which the strengths of the 
potentials of the two surfaces, denoted above by $v_0$ differ. Our results 
for one set of values of $\theta$ and $\phi$ are given in the last two 
columns of table 1. Column 3 gives the maximum amount that $v_0$ for surf1 
can be increased and still get instabilities and column 4 gives the maximum 
amount that $v_0$ can be increased for surf2 and still get instabilities. (There 
is an assymmetry between the surfaces because the angles beteen the direction 
of sliding and the axes of the two surfaces differ.) We find that 
once the strengths of the two surface potentials differ by at most 0.3 percent, 
instabilities are no longer found. This implies that at least at zero 
temperature, there will be no kinetic friction at slow sliding speeds. 
As mentioned earlier, for 
large differences in potential strengths this is not an unexpected result 
because in that case the mobile molecules at the interface are much more 
strongly attached to one surface than the other. This is essentially the 
case of two surfaces in contact at randomly placed points of contact, which 
was considered in Refs. 1-3. There it was found that there is no static 
friction. Since the existence of static and kinetic friction require 
that there be instabilities[4], and since it was shown in Refs. 1-3 that there 
is no static friction, it is also likely that there will be no slow speed 
kinetic friction in this case. The lack of instabilities, and hence slow 
speed kinetic friction, when the potential strengths differ by small amounts, 
comes as a surprise. Since we did find near 
instabilities (i.e., a potential wells bounded by a very low ridge in one 
direction) for case of surfaces whose potential strengths differ by only 
a few percent, the possibility still exists that there will still be 
friction once Boltzmann's constant times the temperature becomes comparable 
to these low potential ridges bounding nearly unstable potential wells. 
An earlier treatment of this problem for two surfaces which consist of a 
random or periodic array of rotationally symmetric Gaussian potential wells[8] 
shows that the minima of the net potential acting on a mobile molecule at 
this model interface will always become unstable as the surfaces slide 
relative to each other. Furthermore, a Gaussian potential well  placed at 
random on one surface, to represent a local defect, can always result in an 
instability, if its depth is greater than $v_0$ [9]. 
Therefore, we concluded that for this model, there 
will always be dry friction for any nonzero temperature. 

\begin{table}
\caption{Results for $\phi=0.1309rad$ and $\theta=0.3927rad$.}
\begin{tabular}{cccc}
b/a&$\Delta E/v_0$&$(\Delta v_1/v_0)\times 10^2$&$(\Delta v_2/v_0)\times 10^2$\\
\tableline
0.0&0.0669&0.285&0.295\\
0.02&0.0571&0.246&0.308\\
0.04&0.0507&0.194&0.304\\
0.06&0.0366&0.132&0.283\\
0.08&0.0162&0.064&0.247\\
0.24&0.0169&0.249&0.064\\
0.26&0.0373&0.287&0.131\\
0.28&0.0510&0.307&0.191\\
0.30&0.0571&0.306&0.241\\
0.32&0.0649&0.280&0.279\\
0.34&0.0993&0.228&0.302\\
0.36&0.1332&0.0&0.0\\
0.38&0.1665&0.0&0.0\\
0.40&0.2667&0.0&0.0\\
0.42&0.2341&0.0&0.0\\
0.44&0.1997&0.0&0.0\\
0.46&0.0565&0.0&0.0\\
0.48&0.1255&0.0&0.0\\
\end{tabular}
\label{table1}
\end{table}

Our conclusion is that although the array of Gaussian potentials treated 
in Ref. 8, which could represent imperfections of the surfaces, appears 
to always exhibit dry friction, the model potential considered in this work, 
which should describe two perfectly periodic surfaces, only exhibits significant 
dry friction when the strengths of the two surface potentials are nearly equal. 
We have also performed molecular dynamic simulations which support the conclusions 
of the procedure used in this work[9].

\acknowledgments

I wish to thank the Department of Energy (Grant DE-FG02-96ER45585).

\end{multicols}{2

\end{document}